%% file: main.tex
\documentclass[aps,prb,twocolumn,superscriptaddress,longbibliography,10pt]{revtex4-2}

\usepackage[utf8]{inputenc}
\usepackage[T1]{fontenc}
\usepackage{amsmath,amssymb,amsfonts}
\usepackage{bm}
\usepackage{graphicx}
\usepackage{hyperref}
\usepackage{siunitx}

\hypersetup{
  colorlinks=true,
  linkcolor=blue,
  citecolor=blue,
  urlcolor=blue
}

\begin{document}

\input{main_body.tex}

\end{document}

%% file: main_body.tex
\title{\texorpdfstring{Proximitized Topological Insulator Charge Island \\ Fabricated via \textit{In Situ} Multi-Angle Stencil Lithography}{Proximitized Topological Insulator Charge Island Fabricated via In Situ Multi-Angle Stencil Lithography}}

\newcommand{\PGIaffil}{PGI-9, Forschungszentrum J\"ulich and JARA J\"ulich-Aachen Research Alliance, 52425 J\"ulich, Germany}

\author{Benedikt Frohn}
 \affiliation{\PGIaffil}

\author{Tobias Schmitt}
 \affiliation{\PGIaffil}

\author{Vanessa Serrano}
 \affiliation{\PGIaffil}

\author{Anne Schmidt}
 \affiliation{\PGIaffil}

\author{Michael Schleenvoigt}
 \affiliation{\PGIaffil}

\author{Albert Hertel}
 \affiliation{\PGIaffil}

\author{Benjamin Bennemann}
 \affiliation{\PGIaffil}

\author{Abdur Rehman Jalil}
 \affiliation{\PGIaffil}

\author{Detlev Gr\"utzmacher}
 \affiliation{\PGIaffil}

\author{Peter Sch\"uffelgen}
 \email{p.schueffelgen@fz-juelich.de}
 \affiliation{\PGIaffil}

\date{April 20, 2026}

\begin{abstract}
Hybrid superconductor--topological insulator (TI) nanostructures constitute a promising material platform for exploring proximity-induced superconductivity in systems with topologically protected surface states. A key obstacle has been the realization of clean and well-controlled superconductor--TI interfaces, as TI surfaces rapidly degrade under ambient conditions. Here, we introduce a fully \textit{in situ}, multi-angle stencil lithography technique that enables the fabrication of proximitized charge islands in TIs. The approach combines selective area growth of (Bi,Sb)$_2$Te$_3$ nanoribbons with angle-controlled deposition of diffusion barriers, superconducting Al, and ultrathin oxide tunnel barriers, allowing scalable fabrication of hybrid nanostructures without post-growth processing. Low-temperature transport measurements reveal robust Coulomb blockade and a pronounced suppression of low-energy conductance, which vanishes with a magnetic field, consistent with proximity-induced superconductivity in the island. These results establish a versatile nanofabrication platform that enables access to previously unexplored TI-based hybrid quantum devices and opens new routes for investigating superconductivity in topological nanostructures.
\end{abstract}

\maketitle

\section{\label{sec:level1}Introduction}

Creating topological superconductivity remains a central objective in condensed matter physics due to its prospective application in fault-tolerant quantum computing \cite{kitaevFaulttolerantQuantumComputation2003}. The essential theoretical building blocks are Majorana zero modes, predicted to emerge in topological systems proximitized by a conventional $s$-wave superconductor \cite{fuSuperconductingProximityEffect2008a, sauGenericNewPlatform2010, lutchynMajoranaFermionsTopological2010b, oregHelicalLiquidsMajorana2010b}. Semiconductor nanowires with strong spin--orbit coupling have therefore been studied extensively, with early experiments reporting signatures consistent with Majorana physics in nanowires \cite{mourikSignaturesMajoranaFermions2012} and superconducting charge islands \cite{higginbothamParityLifetimeBound2015a, albrechtExponentialProtectionZero2016, shenParityTransitionsSuperconducting2018, shenFullParityPhase2021}. These interpretations, however, remain under active debate \cite{valentiniMajoranalikeCoulombSpectroscopy2022}.

\begin{figure}
    \centering
    \includegraphics[width = \linewidth]{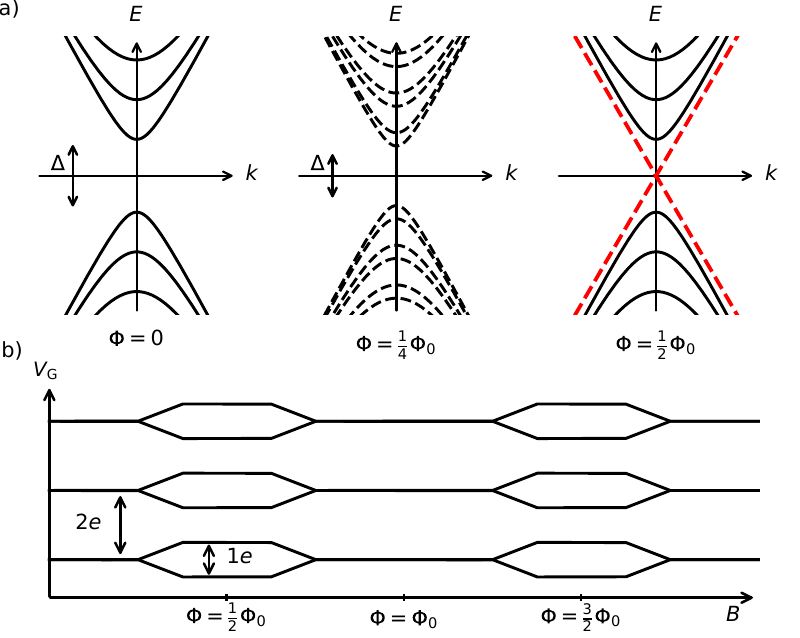}
    \caption{(a) Schematic band structure of a quasi-1D TI nanoribbon as a function of the axial magnetic flux, with a bandgap of $\Delta$. At $\Phi = (n + 1/2)\Phi_0$, the surface-state spectrum (red) re-enters a topological regime. Solid lines indicate doubly degenerate bands, while dashed lines indicate non-degenerate bands.
    (b) Schematic of Coulomb peaks in a proximitized topological-insulator charge island with periodically appearing zero-bias states. When the topology is restored by flux going through the wire [$\Phi = (n + 1/2)\Phi_0$], Majorana zero modes are predicted to appear and thus lift the $2e$ periodicity. The size of the topological regime depends on the exact position of the Fermi level in the TI.}
    \label{fig1}
\end{figure}

\begin{figure*}[t]
    \includegraphics[width = \linewidth]{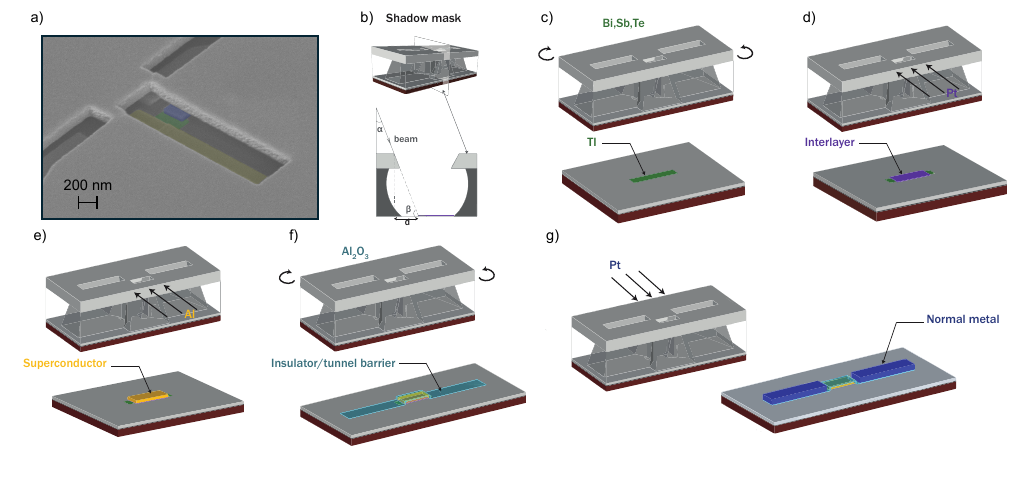}
    \caption{(a) False-color SEM image of a representative device under a tilt of \qty{50}{\degree} showing the TI ribbon (green), superconducting island (yellow), normal contacts (blue), and mask. (b) Pre-patterned shadow mask with its cross section, showing the partially free-standing mask. (c--g) Schematic of the multi-angle \textit{in situ} fabrication process. The TI is grown under rotation in panel c. A Pt diffusion barrier and Al superconductor are deposited from the same angle without rotation in panels d and e. An Al$_2$O$_3$ tunnel barrier is grown under rotation in panel f, followed by deposition of normal Pt contacts from the opposite direction in panel g.}
    \label{fig2}
\end{figure*}

Topological insulators (TIs) provide an alternative platform in which proximity-induced superconductivity is expected to give rise to topological phases \cite{cookStabilityMajoranaFermions2012}. A key conceptual advantage of TIs is that the topological regime can be realized periodically over a wide energy range by adjusting the chemical potential, the magnetic flux along the wire, or both \cite{cookStabilityMajoranaFermions2012, cookMajoranaFermionsTopologicalinsulator2011, heffelsRobustFragileMajorana2023a}. Spectroscopy on quasi-one-dimensional TI nanowires offers a controlled setting, as their topological properties are governed by the magnetic flux threading the wire. In particular, for a wire with a uniform cross section, at half-integer flux quanta, $\Phi = (n + 1/2)\Phi_0$, the surface-state spectrum is expected to re-enter a topologically non-trivial regime (Fig.~\ref{fig1}a), while trivial phases occur at integer flux quanta \cite{heffelsRobustFragileMajorana2023a}. This alternating sequence provides a robust experimental fingerprint. Despite these favorable properties, spectroscopic studies of proximitized TIs remain comparatively scarce \cite{finckPhaseCoherenceAndreev2014, sunMajoranaZeroMode2016, wiedenmannTransportSpectroscopyInduced2017a, stehnoConductionSpectroscopyProximity2017}, with tunnel spectroscopy only recently showing significant progress \cite{schluckRobustGapClosing2024, fengLongrangeCrossedAndreev2025a}.

Charge island hybrid devices constitute a complementary approach to tunneling spectroscopy, as they allow access to charging effects and parity-dependent transport \cite{albrechtExponentialProtectionZero2016}. Beyond spectroscopy, control over fermion parity is a key requirement for the operation and readout of Majorana-based qubits \cite{fuElectronTeleportationMajorana2010a, vanheckCoulombassistedBraidingMajorana2012, aasenMilestonesMajoranaBasedQuantum2016, pluggeMajoranaBoxQubits2017}. While hybrid charge island devices are well-established in semiconductor--superconductor hybrids, their realization in TIs has been limited to a few non-proximitized examples \cite{choTopologicalInsulatorQuantum2012a, jingSingleElectronTransistorMade2019a, witmansQuantumTransportSnTe2025b, atanovTopologicalInsulatorSingleelectron2026}. In a superconducting charge island, the charging periodicity is governed by the competition between the charging energy and the energy of the lowest-lying odd-parity excitation. If odd-parity states remain energetically inaccessible, transport occurs through Cooper pairs and a $2e$ charging periodicity is expected. In contrast, low-energy subgap states reduce the energetic cost of odd-parity charge states and can lead to even--odd or $1e$-periodic transport \cite{shenParityTransitionsSuperconducting2018}. In a topological regime, a zero-energy state is expected to provide such a low-energy odd-parity state, resulting in a flux-dependent modification of the charging periodicity, as schematically illustrated in Fig.~\ref{fig1}b. To the best of our knowledge, a proximitized TI charge island has not been demonstrated so far. This absence is largely attributable to two material challenges. First, unlike semiconductors, TIs cannot be fully depleted into a bulk band gap by electrostatic gating due to the presence of metallic surface states. Second, rapid surface oxidation \cite{kongRapidSurfaceOxidation2011} degrades interface transparency, complicating the formation of a high-transparency superconductor--TI interface.

In this work, we report transport measurements on a proximitized TI charge island fabricated using an advanced \textit{in situ} multi-angle growth technique that directly addresses these challenges. The process enables the formation of pristine interfaces for proximity-induced superconductivity and the realization of well-defined tunneling barriers. We observe robust Coulomb blockade over a wide gate voltage range and a pronounced suppression of low-energy conductance consistent with an induced superconducting gap in the island.

\section{\label{sec:Fab}Fabrication}

The fabrication process builds on selective area growth (SAG) \cite{jalilSelectiveAreaEpitaxy2023} combined with shadow evaporation, extending the approach introduced in Ref.~\cite{schuffelgenSelectiveAreaGrowth2019}. A detailed description of the mask fabrication is provided in the Supporting Information. A lower mask defines the selective growth region for the TI, while an upper, partially free-standing mask enables shadowed deposition of multiple materials. With the control of the evaporation angles and sample rotation, spatial offsets between different layers are achieved without any post-growth processing, allowing the realization of hybrid nanostructures with integrated tunneling electrodes in a single growth sequence.

\begin{figure*}[t]
    \includegraphics[width = \linewidth]{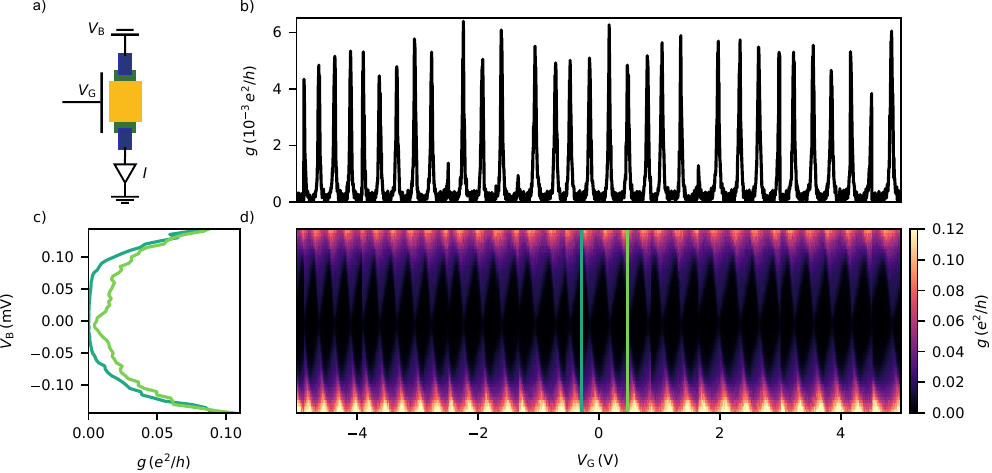}
    \caption{(a) Measurement schematic of the proximitized charge island. A bias voltage is applied to one contact, while the current is measured on the other one. (b) Differential conductance $g$ as a function of back-gate voltage at zero bias, showing periodic Coulomb blockade peaks. (c) Differential conductance versus bias voltage, for two line cuts at different gate voltages. In turquoise is a line cut through complete blockade at low bias, while in green, the Coulomb diamond is exactly closed. (d) Coulomb diamonds measured as a function of bias and gate voltage, demonstrating a well-defined charging regime. Line cuts from panel c are marked in the same colors.}
    \label{fig3}
\end{figure*}

Figure~\ref{fig2} illustrates the growth procedure. Figure~\ref{fig2}a shows a false-color scanning electron micrograph of a representative device, highlighting the TI nanoribbon (green), superconducting island (yellow), normal electrodes (blue), and mask/substrate (gray). The unusually large central opening was chosen to improve imaging of the underlying structure. The double-mask stack consists of a lower \qty{5}{\nano\meter} SiO$_2$ and \qty{20}{\nano\meter} Si$_3$N$_4$ layer defining the trenches for SAG and an upper \qty{300}{\nano\meter} SiO$_2$ and \qty{100}{\nano\meter} Si$_3$N$_4$ layer forming the shadow mask (see Fig.~\ref{fig2}b). The trenches are patterned by lithography in Si$_3$N$_4$, followed by wet etching of SiO$_2$ to create the partially free-standing geometry.

Following the recipe in Ref.~\cite{schuffelgenSelectiveAreaGrowth2019}, the TI nanoribbon is grown first under rotation, confined to the rectangular selective area, to a thickness of \qty{18}{\nano\meter}, width of \qty{200}{\nano\meter}, and length of \qty{1}{\micro\meter} with a composition of Bi$_{2-x}$Sb$_x$Te$_3$, with $x = 1.68 \pm 0.04$ (see Fig.~\ref{fig2}c). To suppress interdiffusion between the TI and the superconductor, a \qty{3}{\nano\meter} Pt diffusion barrier is deposited \cite{schmittAnomalousTemperatureDependence2022a}, followed by a \qty{20}{\nano\meter} Al layer acting as the proximitizing superconductor (Fig.~\ref{fig2}d and e). Both are deposited from the same direction without rotation and defined by the middle hole in the mask. An oscillatory motion of the substrate holder relative to the material source during deposition can be used to adjust the shadow and increase coverage \cite{schmittIntegrationTopologicalInsulator2022}. A stoichiometric Al$_2$O$_3$ tunnel barrier is subsequently grown under rotation to cap the exposed ends of the TI ribbon (Fig.~\ref{fig2}f). Although the nominal thickness is \qty{5}{\nano\meter}, sequential shadowing by the upper mask during rotation reduces the effective barrier thickness to approximately \qty{1}{\nano\meter}. Finally, \qty{40}{\nano\meter} thick Pt normal electrodes are evaporated from the opposite angle with respect to the superconductor deposition, shown in Fig.~\ref{fig2}g, and the entire structure is capped with Al$_2$O$_3$ under rotation to protect the final device from degradation at ambient conditions (not shown).

This fully \textit{in situ} process offers several advantages. It prevents oxidation of the TI surface prior to superconductor deposition \cite{ngabonzizaSituSpectroscopyIntrinsic2015}, avoids damage and contamination associated with post-growth etching \cite{schuffelgenSelectiveAreaGrowth2019}, and enables fabrication of many devices on the same chip. Moreover, despite the exponential dependence of tunnel rates on barrier thickness, the resulting barriers are reproducible across devices grown in the same run, enabling more complex multi-barrier architectures.

\section{Transport Measurements}

\begin{figure*}[t]
    \includegraphics[width = \linewidth]{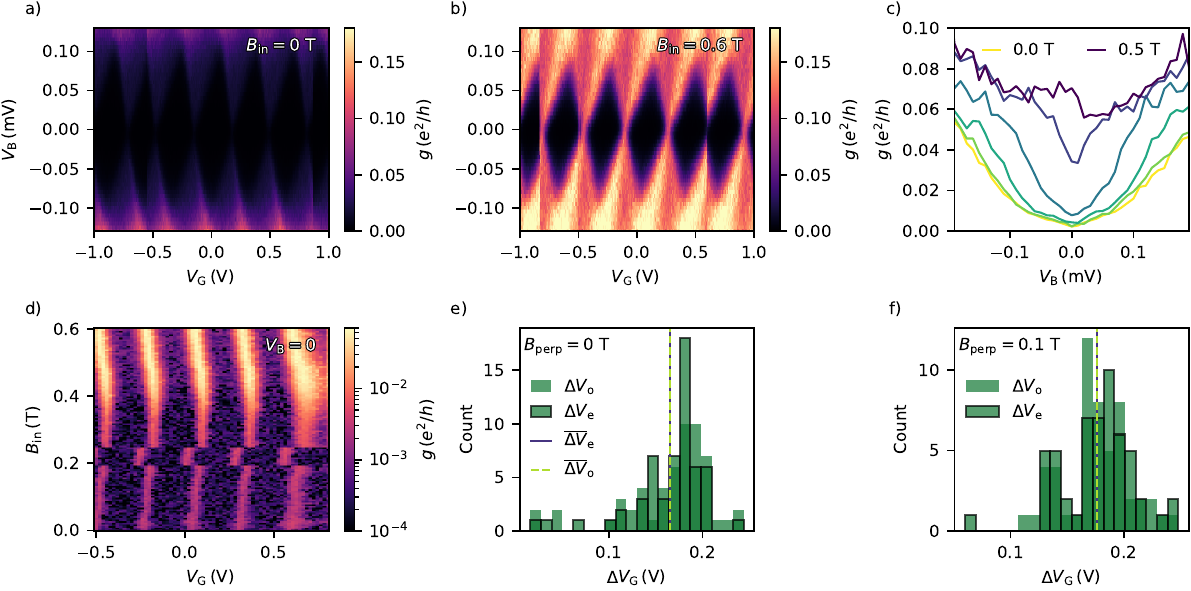}
    \caption{(a and b) Coulomb diamonds at zero in-plane magnetic field and \qty{0.6}{\tesla}. A pronounced suppression of conductance outside the diamonds is visible at zero field, which gets lifted at high field. (c) Line cuts through the closing points of the Coulomb diamonds at magnetic fields spaced by \qty{0.1}{\tesla}, highlighting the magnetic-field-dependent conductance suppression associated with proximity-induced superconductivity. (d) Gate voltage versus magnetic field at zero bias. The gate periodicity stays constant, but the conductance increases once the superconductivity is diminished. (e and f) Histograms of even--odd $\Delta V_\mathrm{e}$ and odd--even $\Delta V_\mathrm{o}$ peak spacings at zero field and high out-of-plane field. The average spacing is marked by a solid blue and dashed green line.}
    \label{fig4}
\end{figure*}

Electronic transport measurements are performed in a dilution refrigerator with a base temperature of \qty{10}{\milli\kelvin} and an estimated electron temperature of \qty{50}{\milli\kelvin}. Standard low-frequency lock-in techniques are employed to measure the differential conductance with an excitation voltage of \qty{7}{\micro\volt} for the data presented in Figs.~\ref{fig3} and \ref{fig4}a and b, while \qty{10}{\micro\volt} was used everywhere else including the Supporting Information. A schematic of the measurement configuration is shown in Fig.~\ref{fig3}a. A voltage bias is applied to one normal electrode, while the other electrode is grounded, and the resulting current is measured. A global back gate is realized by applying a voltage to the metallic back plate of the chip carrier, with the chip made of high-resistivity Si constituting the required dielectric.

Figure~\ref{fig3}b shows periodic conductance peaks as a function of gate voltage at zero bias, characteristic of Coulomb blockade. Bias spectroscopy at a fixed gate voltage (Fig.~\ref{fig3}c) reveals suppressed conductance at $\num{-0.05} < V_\mathrm{B} < \qty{0.08}{\milli\volt}$. The two line cuts are taken at $V_\mathrm{G} = \num{-0.28}$ and \qty{0.48}{\volt} for turquoise and green, respectively. Combining sweeps of both bias and gate voltage yields well-defined Coulomb diamonds over a wide gate range (Fig.~\ref{fig3}d), unambiguously demonstrating quantized charge transport through the island. The line cuts in Fig.~\ref{fig3}c are chosen such that they capture a blockaded configuration (turquoise) and an open one (green). The slight asymmetry of the conductance suppression around zero bias in Fig.~\ref{fig3}c is attributed to the biasing scheme and asymmetry of the electrical contacts.

From the diamond size, we extract a charging energy of \qty{95}{\micro\electronvolt} and a gate lever arm of approximately $4\times10^{-4}$. The charging energy remains constant over a wide gate range, indicating a highly occupied regime. The extracted charging energy is lower than expected from electrostatic simulations ($\approx\qty{360}{\micro\electronvolt}$), likely due to enhanced screening by the surrounding metallic mask structure, which increases the total island capacitance. Notably, a smaller effective island would yield a larger charging energy; the observed reduction therefore indicates that the island extends across the full TI trench and superconducting island rather than being confined to a subset of it.

To probe the influence of the superconducting proximity effect, Fig.~\ref{fig4}a and b show Coulomb diamonds at zero field and an in-plane field along the TI wire $B_\mathrm{in} = \qty{0.6}{\tesla}$. At zero field, the differential conductance outside the blockade regime is strongly suppressed at low bias, consistent with a proximity-induced gap in the density of states. At high field, where superconductivity is suppressed, this additional reduction in conductance disappears. Figure~\ref{fig4}c presents line cuts (positions marked in Fig.~\ref{fig:SI_Bin}) taken between two Coulomb diamonds, directly illustrating the closing of the induced gap.

Independent evidence for the formation of an induced superconducting gap is obtained from tunnel spectroscopy of a longer proximitized TI nanoribbon fabricated on the same chip and therefore under the same growth conditions (Fig.~\ref{fig:supp_gap_reference}). In this device, the superconducting contact is grounded and spectroscopy at one end of the nanoribbon reveals an induced gap of approximately $\Delta^* \approx \qty{200}{\micro\electronvolt}$. While the magnitude of the induced gap in this extended device does not directly determine the gap of the charge island, the measurement demonstrates that the same fabrication process and growth conditions produce proximity-induced superconductivity in the TI nanoribbon.

In Fig.~\ref{fig4}d, we map the conductance as a function of the gate voltage and magnetic field at near-zero bias. While the Coulomb peak conductance increases once superconductivity is suppressed ($B_\mathrm{c,in} \approx \qty{0.4}{\tesla}$), the peak spacing remains unchanged across the entire field range. No transition from $1e$ to $2e$ periodicity is observed. The sudden jump near $B_\mathrm{in} = \qty{0.2}{\tesla}$ is due to charge rearrangements.

Statistical analysis of 108 consecutive Coulomb peaks is shown in Fig.~\ref{fig4}e and f, where we compare odd-to-even $\Delta V_\mathrm{o}$ and even-to-odd $\Delta V_\mathrm{e}$ spacings at zero field and a high out-of-plane field ($B_\mathrm{perp} = \qty{0.1}{\tesla}$), at which superconductivity in the device is suppressed. Within experimental uncertainty, no statistically significant difference between the average of the two spacings $\Delta\overline{V}_\mathrm{o}$ and $\Delta\overline{V}_\mathrm{e}$ is found. We also do not find an indication that the distribution of the spacings changes in a systematic manner when the superconductivity is suppressed. The broad distribution of spacings is attributed to charge rearrangements in the electrostatic environment, likely dominated by trapped charges in the surrounding oxides.

The absence of $2e$ periodicity is consistent with the presence of a soft gap in the proximitized island. Importantly, the relevant energy scale for the island parity is the energy $E_0$ of the lowest-lying odd-parity subgap excitation rather than the magnitude of the superconducting gap $\Delta^*$ itself \cite{shenParityTransitionsSuperconducting2018}. If sufficiently low-energy subgap states are present, such that $E_0$ is small compared to the charging energy, odd-parity charge states can become energetically accessible and single-electron transport can occur even in the presence of a finite superconducting gap. In our device, not all electronic states are necessarily proximitized equally strongly, and weakly or non-proximitized states may therefore remain at low energies and prevent $2e$ periodicity. Contributions from the diffusion barrier cannot be excluded and motivate further optimization of material combinations and interface engineering. In addition, the use of normal-conducting leads may further suppress $2e$ periodicity in the island. The leads can be a source of quasiparticles, which are known to prevent parity preservation \cite{aumentadoNonequilibriumQuasiparticles$2e$2004a}. We deliberately avoided superconducting leads in order to unambiguously attribute the observed superconductivity to the proximity effect of the island.

\section{Conclusion \& Outlook}

In conclusion, we demonstrate a proximitized topological insulator charge island realized using a fully \textit{in situ} multi-angle fabrication technique. The approach yields clean superconductor--TI interfaces, reproducible tunnel barriers, and scalable device fabrication. Low-temperature transport measurements reveal stable Coulomb blockade and a clear proximity-induced superconducting gap, establishing a new experimental platform for hybrid TI--superconductor systems. While no charge-parity preservation is observed, the results identify key material challenges and provide a foundation for future studies. The fabrication scheme is readily adaptable to tunnel spectroscopy of extended TI nanowires and more complex device architectures. Overall, this work advances experimental control in TI--superconductor hybrids and supports a more systematic exploration of topological superconductivity and Majorana physics in 3D topological insulators.

\section*{Data availability}
The data and code used in the generation of the main and supplementary figures are available on Zenodo at \url{https://doi.org/10.5281/zenodo.23214639}.

\begin{acknowledgments}
The authors acknowledge Stefan Trellenkamp and Florian Lentz for their assistance with electron-beam lithography and related processes.
This work has been supported financially by the German Federal Ministry of Research, Technology and Space (BMFTR) via Quantum Future Project ``MajoranaChips'' (Grant 13N15264), and the Deutsche Forschungsgemeinschaft (DFG, German Research Foundation) under Germany's Excellence Strategy---Cluster of Excellence Matter and Light for Quantum Computing (ML4Q) EXC 2004/1-390534769.
\end{acknowledgments}

\bibliography{References}

\clearpage
\onecolumngrid
\suppressfloats[t]

\setcounter{section}{0}
\setcounter{subsection}{0}
\setcounter{equation}{0}
\setcounter{figure}{0}
\setcounter{table}{0}

\renewcommand{\thefigure}{S\arabic{figure}}
\renewcommand{\thetable}{S\arabic{table}}

\begin{center}
    {\Large\bfseries Supporting Information:\par}
    \vspace{0.5em}
    {\large\bfseries Proximitized Topological Insulator Charge Island \\ Fabricated via \textit{In Situ} Multi-Angle Stencil Lithography\par}
    \vspace{1em}
    {Benedikt Frohn, Tobias Schmitt, Vanessa Serrano, Anne Schmidt, Michael Schleenvoigt, Albert Hertel, Benjamin Bennemann, Abdur Rehman Jalil, Detlev Gr\"utzmacher, and Peter Sch\"uffelgen$^{*}$\par}
    \vspace{0.5em}
    {\small PGI-9, Forschungszentrum J\"ulich and JARA J\"ulich-Aachen Research Alliance, 52425 J\"ulich, Germany\par}
    \vspace{0.3em}
    {\small $^{*}$Email: p.schueffelgen@fz-juelich.de\par}
\end{center}

\section*{Electrostatic properties}

The charging energy of the superconducting island was evaluated using a three-dimensional electrostatic model implemented in COMSOL Multiphysics. The simulated structure consisted of a topological-insulator (BST) nanowire contacted by superconducting metallic islands and leads. The BST channel was modeled with a length of \qty{1}{\micro\meter} and a width of \qty{200}{\nano\meter}. The superconducting island had a length of \qty{800}{\nano\meter} and a width of \qty{250}{\nano\meter}, while the contact electrodes were modeled with a width of \qty{300}{\nano\meter}. To reduce computational complexity, the lithographic mask used during fabrication was omitted from the simulation geometry. This simplification is expected to underestimate the total capacitance and, consequently, overestimate the charging energy relative to the fabricated device. All material parameters were defined for the few-kelvin regime. Superconducting aluminum was represented by a conductivity of \qty{1e9}{\siemens\per\meter} to approximate ideal metallic behavior within the electrostatic model. Intrinsic silicon was treated as an insulating substrate with a conductivity of \qty{1e-12}{\siemens\per\meter}, consistent with carrier freeze-out at cryogenic temperatures. The material parameters used in the simulations are summarized in Table~\ref{tab:material_properties}.

\begin{table}[htbp]
\centering
\caption{Material parameters used in the COMSOL electrostatic simulations. Values for Pt correspond to cryogenic transport properties in the few-kelvin regime. The conductivity of Al was chosen to approximate superconducting behavior, while the conductivity of intrinsic Si represents the insulating limit arising from carrier freeze-out at low temperature.}
\label{tab:material_properties}
\begin{tabular}{lccc}
\hline\hline
Material &
$\varepsilon_r$ &
$\rho$ (\si{\ohm\meter}) &
$\sigma$ (\si{\siemens\per\meter}) \\
\hline
Vacuum & 1 & -- & 0 \\
Si & 11.7 & -- & \num{1e-12} \\
Al & 1 & -- & \num{1e9} \\
Pt & 1 & \num{2.6e-11} & -- \\
Al$_2$O$_3$ & 9.7 & -- & -- \\
\hline\hline
\end{tabular}
\end{table}

The gate lever arm, $\alpha_{\mathrm{G}}$, quantifies the coupling between the gate voltage and the electrochemical potential of the island. It is extracted from the slopes of the Coulomb-diamond edges. The two slopes, $m_1$ and $m_2$, are defined using the convention $m_i=\mathrm{d}V_{\mathrm{G}}/\mathrm{d}V_{\mathrm{B}}$. For the asymmetric biasing configuration used here, the inverse gate lever arm is given by the absolute difference between the two slopes,
\begin{equation}
|\Delta m|
= |m_1-m_2|
= \frac{1}{\alpha_{\mathrm{G}}}.
\label{eq:LeverarmAsym}
\end{equation}

\section*{Device fabrication and stencil-mask preparation}

The devices were fabricated on high-resistivity Si(111) substrates purchased from CrysTec GmbH, with a room-temperature resistivity of 5000--10000~\si{\ohm\centi\meter}. Although the substrates are nominally weakly conducting at room temperature, they become effectively insulating at cryogenic temperatures. The substrate type is equivalent to that used in Ref.~\cite{schuffelgenSelectiveAreaGrowth2019}, although the substrates may originate from a different manufacturing batch.

The double-mask structure was fabricated in two sequential steps. First, the lower selective-area-growth mask was prepared following the process described in Ref.~\cite{jalilSelectiveAreaEpitaxy2023}. This mask consists of a thin SiO$_2$/Si$_3$N$_4$ stack with a total thickness of approximately 25--\qty{30}{\nano\meter}. Openings in this mask define the exposed Si(111) trenches in which the quasi-one-dimensional topological-insulator nanostructures are grown.

Second, an upper stencil/shadow mask was fabricated on top of the selective-area-growth mask following the approach introduced in Ref.~\cite{schuffelgenSelectiveAreaGrowth2019}. The upper mask consists of \qty{300}{\nano\meter} of SiO$_2$ and \qty{100}{\nano\meter} of Si$_3$N$_4$. The vertical separation between the lower growth plane and the upper stencil mask, together with the designed lateral openings, defines the shadow geometry used for the angle-selective deposition of the superconducting and normal-metal layers.

All nanoscale openings were patterned by electron-beam lithography. The Si$_3$N$_4$ layers were structured by reactive-ion etching using CHF$_3$/O$_2$ chemistry. Immediately before loading the samples into the molecular-beam-epitaxy system, the SiO$_2$ layers were selectively etched using wet HF. This etching step removes the oxide from the bottom of the selective-area-growth trenches, thereby exposing the Si(111) surface for topological-insulator growth. Simultaneously, it underetches the sacrificial SiO$_2$ layer of the upper mask and releases the free-standing Si$_3$N$_4$ stencil bridges.

\begin{figure}[tbp]
    \centering
    \includegraphics{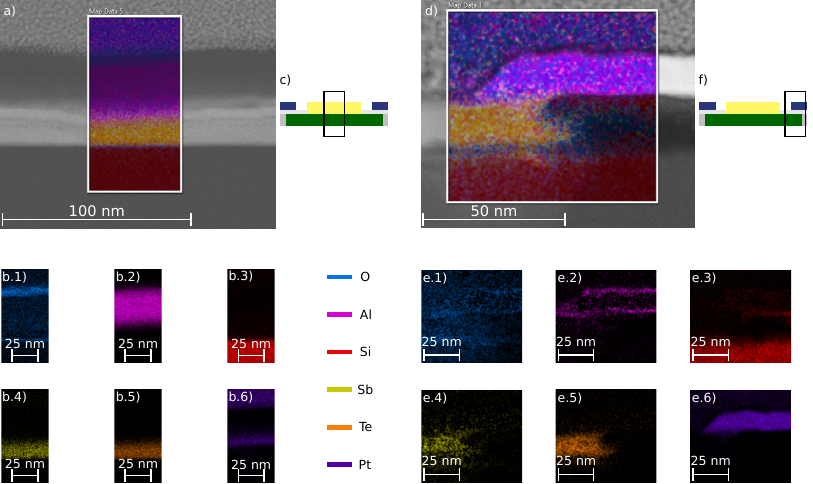}
    \caption{Transmission electron microscopy (TEM) images and elemental composition maps of a device fabricated using the same process as the device studied in transport measurements. Two cross-sectional positions are shown: (a) the Al--Pt--TI interface and (d) the Pt--Al$_2$O$_3$--TI interface. Their locations within the device are indicated schematically in (c) and (f), respectively. Panels (b.1--b.6) and (e.1--e.6) show the element-specific signals for the corresponding regions.}
    \label{fig:TEM}
\end{figure}

After loading into the growth chamber, the samples were annealed at \qty{700}{\degreeCelsius} for \qty{20}{\minute}. This annealing step removes the hydrogen termination produced by the HF treatment and exposes reactive Si dangling bonds. During the initial growth sequence, the surface is exposed to Te, which passivates the dangling bonds and forms a suitable interface for the subsequent van der Waals epitaxy of the BST nanoribbon within the predefined selective-area-growth trenches. Full process development of the lower selective-area-growth mask and the upper stencil-mask process is provided in Ref.~\cite{jalilSelectiveAreaEpitaxy2023} and Ref.~\cite{schuffelgenSelectiveAreaGrowth2019}, respectively.

The resulting material stack is illustrated in Fig.~\ref{fig:TEM} using cross-sectional transmission electron microscopy and elemental composition analysis performed on a device fabricated with the same process as the measured devices. The images show the two relevant interface regions, namely the Al--Pt--TI interface and the Pt--Al$_2$O$_3$--TI interface. The corresponding elemental maps confirm the spatial arrangement of the deposited layers and demonstrate the successful formation of the intended heterostructure following selective area growth and angle-dependent metal deposition. As the investigated lamella was extracted from a nominally similar device rather than from the device used for transport measurements, the microscopy results are representative of the fabrication process and material stack.

\section*{Additional magnetic field measurements}

\begin{figure}[htbp]
  \centering
  \includegraphics{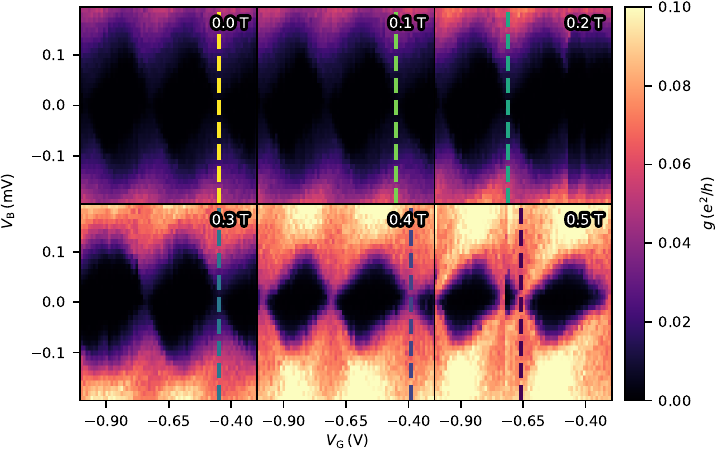}
  \caption{Coulomb-diamond measurements at different in-plane magnetic fields. The magnetic field applied in each measurement is indicated in the upper-right corner of the corresponding panel. With increasing magnetic field, the conductance outside the Coulomb-blockaded regions increases. The dashed lines indicate the positions of the line cuts shown in Fig.~\ref{fig4}c of the main text.}
  \label{fig:SI_Bin}
\end{figure}

\begin{figure}[htbp]
    \centering
    \includegraphics{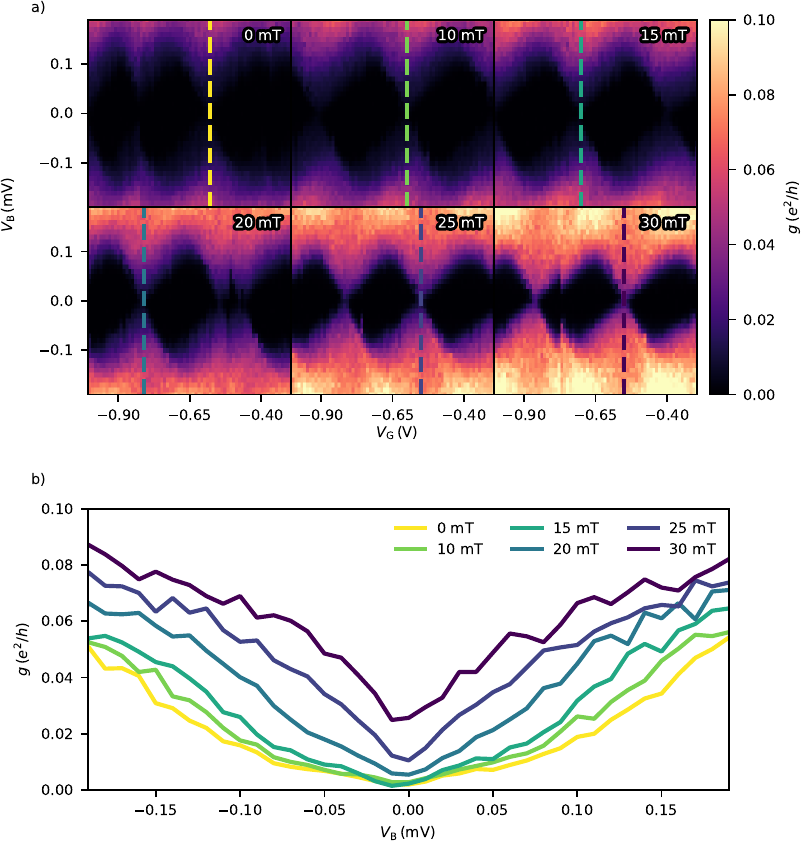}
    \caption{(a) Coulomb-diamond measurements at different out-of-plane magnetic fields. The magnetic field applied in each measurement is indicated in the upper-right corner of the corresponding panel. With increasing magnetic field, the conductance outside the Coulomb-blockaded regions increases. (b) Line cuts at different magnetic fields, taken at the positions marked by the dashed lines in (a).}
    \label{fig:SI_Bout}
\end{figure}

The magnetic-field dependence of the induced superconducting gap is examined for both in-plane and out-of-plane field orientations in Figs.~\ref{fig:SI_Bin} and \ref{fig:SI_Bout}, respectively. The magnetic field applied in each Coulomb-diamond measurement is indicated in the upper-right corner of the corresponding panel. In both configurations, increasing the magnetic field enhances the conductance outside the Coulomb-blockaded regions, consistent with a progressive suppression of the induced gap. For the in-plane field measurements, the dashed lines in Fig.~\ref{fig:SI_Bin} indicate the positions of the line cuts presented in Fig.~\ref{fig4}c of the main text. The corresponding out-of-plane field line cuts are shown together with the Coulomb-diamond measurements in Fig.~\ref{fig:SI_Bout}.

\section*{Independent characterization of proximity-induced superconductivity}

To independently characterize the superconducting proximity effect produced by our fabrication process, we perform tunnel spectroscopy on a longer proximitized TI nanoribbon fabricated on the same chip as the charge-island devices. The device therefore experienced the same TI growth, interface preparation, diffusion-barrier deposition, and superconducting deposition. In contrast to the charge-island geometry, the superconducting electrode is grounded and the differential conductance is measured at one end of the proximitized nanoribbon.

\begin{figure}[htbp]
    \centering
    \includegraphics{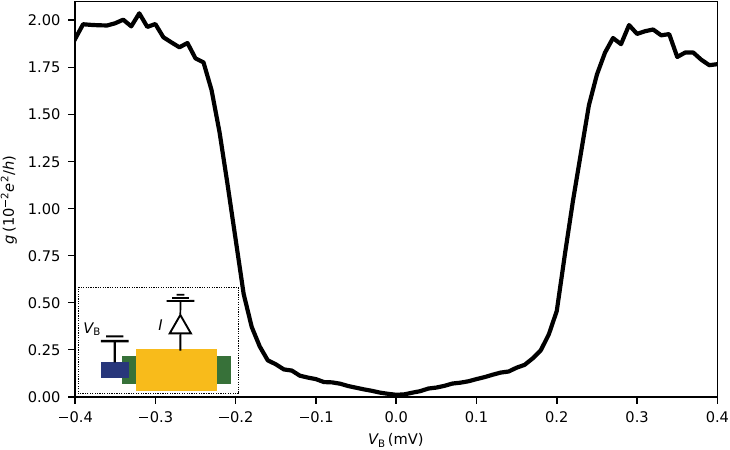}
    \caption{Independent characterization of proximity-induced superconductivity in a TI nanoribbon fabricated on the same chip as the charge-island devices. The differential conductance as a function of bias voltage $V_\mathrm{B}$ shows a pronounced suppression around zero bias, corresponding to an induced superconducting gap of approximately $\Delta^* \approx \qty{200}{\micro\electronvolt}$. Inset: Measurement schematic. The bias voltage $V_\mathrm{B}$ is applied to the normal-metal contact (blue) at one end of the TI nanoribbon (green), and the current $I$ is measured through the grounded superconducting electrode (yellow).}
    \label{fig:supp_gap_reference}
\end{figure}

Figure~\ref{fig:supp_gap_reference} shows the differential conductance as a function of bias voltage $V_\mathrm{B}$. A pronounced suppression of conductance is observed around zero bias, corresponding to an induced superconducting gap of approximately $\Delta^* \approx \qty{200}{\micro\electronvolt}$.

The observation of a well-defined induced gap in this device demonstrates that the fabrication process produces proximity-induced superconductivity in the TI under the same growth conditions as those used for the charge island. The value of $\Delta^*$ obtained here should, however, not be interpreted as a quantitative measurement of the induced gap in the charge island, since the strength of the proximity effect can depend on the individual device geometry, interface coupling, and electronic states involved.

